\documentclass[aps,pra,twocolumn,amsmath,amssymb,superscriptaddress,longbibliography]{revtex4-1}
\usepackage[english]{babel}
\usepackage{amssymb}
\usepackage{dcolumn}
\usepackage{bm}
\usepackage{graphicx}
\usepackage{amsmath}
\usepackage{graphicx}        % standard LaTeX graphics tool
\graphicspath{{pict/}{}}

\usepackage{dcolumn}%%%%%new
\usepackage{bm}
\usepackage[pdfstartview=FitH, CJKbookmarks=true, bookmarksnumbered=true, bookmarksopen=true, colorlinks=true, pdfborder=001, citecolor=blue, linkcolor=blue, urlcolor=blue, linktocpage=true] {hyperref}

\begin{document}

\title{Floquet-surface bound states in the continuum in a resonantly driven 1D tilted defect-free lattice
}

\author{Bo Zhu}
\affiliation{Institute of Mathematics and Physics, Central South University of Forestry and Technology, Changsha 410004, China}
\affiliation{Guangdong Provincial Key Laboratory of Quantum Metrology and Sensing $\&$ School of Physics and Astronomy, Sun Yat-Sen University (Zhuhai Campus), Zhuhai 519082, China}
\author{Yongguan Ke}
\affiliation{Guangdong Provincial Key Laboratory of Quantum Metrology and Sensing $\&$ School of Physics and Astronomy, Sun Yat-Sen University (Zhuhai Campus), Zhuhai 519082, China}
\affiliation{Nonlinear Physics Centre, Research School of Physics, The Australian National University, Canberra ACT 2601, Australia}
\author{Wenjie Liu}
\affiliation{Guangdong Provincial Key Laboratory of Quantum Metrology and Sensing $\&$ School of Physics and Astronomy, Sun Yat-Sen University (Zhuhai Campus), Zhuhai 519082, China}
\author{Zheng Zhou}
\affiliation{Department of Physics, Hunan Institute of Technology, Hengyang 421002, China}
\author{Honghua Zhong}
\altaffiliation{Corresponding author: hhzhong115@163.com.}
\affiliation{Institute of Mathematics and Physics, Central South University of Forestry and Technology, Changsha 410004, China}

\date{\today}

\begin{abstract}

We study the Floquet-surface bound states embedded in the continuum
(BICs) and bound states out the continuum (BOCs)
in a resonantly driven 1D tilted defect-free lattice.
In contrast to fragile single-particle BICs assisted by specially tailored potentials, we find that Floquet-surface BICs, stable against structural perturbations, can exist in a wide range of parameter space.
By using a multiple-time-scale asymptotic analysis in the high-frequency limit, the appearance of Floquet-surface bound states can be analytically explained by
effective Tamm-type defects at boundaries induced by the resonance between the periodic driving and tilt.
The phase boundary of existing Floquet-surface states is also analytically given.
Based on the repulsion effect of surface states, we propose to
detect transition points and measure the number of Floquet-surface bound states by quantum walk.
Our work opens a new door to experimental realization of BICs in quantum system.

\end{abstract}

\maketitle

\section{Introduction\label{Sec1}}
Surface bound states in the continuum (BICs),
localized interface waves with energy penetrating into a continuous spectrum of radiative waves,
have attracted much attention in several physical fields
ranging form condensed matter physics to
optics ~\cite{hsu2016, kosh2019,yang2013,mur2014,sab2015,molina2012,longhi2014,della2013,gallo2015}.
Their unique properties have
led to numerous applications, including lasers, sensors,
filters, low-loss fibres and Raman spectroscopy \cite{hsu2016, kosh2019}.
Surface BICs generally are regarded as fragile
states that usually decay into resonance surface states when the system parameters are slightly perturbed,
and thus can exist in  only a few special systems. Recently, theory works have suggested surface BICs in a one-dimensional (1D) lattice with
tailored potentials~\cite{molina2012,longhi2014,della2013,gallo2015}.
Furthermore, surface BICs with algebraic \cite{corr2012} or
compact \cite{weiman2013} localization have been demonstrated in
experiments by using photonic structures that allow robust control of parameters.
These types of surface BICs manifest themselves through inverse construction achieved
by engineering the potential or the hopping rate.
Therefore, such
previous studies have been limited to consider static (i.e.,
undriven) lattices, and then surface BICs tuned from a single resonance is lacking.
In particular, the realization of a quantum surface BICs still
remains a challenge.

Recently, the concept of Floquet
BICs has been introduced in a periodically driven 1D
tight-binding defective lattice \cite{santander2013, longhi2219,sen2017}. By tailing inhomogeneous hopping rates and applying external sinusoidal driving, Floquet BICs appear as a
result of selective destruction of tunneling~\cite{longhi2219}.
As happened in other contexts, moving to the many-particle framework,
two-particle Floquet BICs have been predicted to exist in
defect-free Hubbard lattices, either in the bulk~\cite{zhong2017} or
at the surface~\cite{valle2014}. It has been shown that in the high-frequency
limit,
the external periodic driving can induce an virtual surface defect
in the defect-free semilattice, and thus results in two-particle Floquet-surface BICs~\cite{valle2014}.
However, strong particle interaction and bichromatic driving play a key role in
the formation of Floquet-surface BICs, otherwise no Floquet-surface BICs
were observed without particle interaction~\cite{valle2014}. Therefore, a natural question arises: Can single-particle Floquet-surface BICs
be realized using 1D defect-free lattice?

In this paper, we show that
single-particle Floquet-surface BICs appear in a resonantly driven 1D tilted defect-free lattice, which can be readily realized in cold atom systems where resonantly modulated tilted lattices have applied to produce artificial magnetic fields~\cite{aide2011,aide2013,aides2013,aide2015,miya2013}
or to control tunneling dynamics~\cite{Sias2008, tan2014, Goldman2015, tan2016}.
We find pairs of Floquet-surface BICs and BOCs in a wide range of parameter space  are immune to perturbations of system parameters, in stark contrast to static surface BICs that require the lattice possessing intrinsic
surface impurity~\cite{molina2012,longhi2014,della2013}, disorder~\cite{molina2012}
or inhomogeneous hopping rates~\cite{corr2012}.
Based on the repulsion effect related to the localization properties of surface states~\cite{wang2017quantum}, we propose to use quantum walk for
detecting transition points and measuring  the number of Floquet-surface states.
We have successfully explain the underlying mechanism  for the formation of Floquet-surface states via a multiple-time-scale asymptotic analysis, that is, the resonance between the periodic driving and tilt can induce effective Tamm-type defects at boundaries of the lattice,
and thus results in the appearance of Floquet-surface states.
The boundary of existing Floquet-surface states is analytically given, which can be tuned  by the coupling strength and the driving amplitude.
We should emphasize that the previous relative studies focus on the delocalization in the bulk induced by the resonant modulations~\cite{ma2011photon, tan2014}. To the best of our knowledge, this is the first time to show localization at the edges of lattice induced by resonant modulations.

The rest of paper is organized as follows.
In Sec.~\ref{Sec2}, we introduce the model, Floquet BICs and BOCs and their detection via quantum walks.
In Sec.~\ref{Sec3}, we apply multiple-time-scale asymptotic analysis in the high frequency limit to understand the formation of Floquet-surface bound states and parameter boundary for the existence of Floquet-surface bound states.
At last, we give a conclusion in Sec.~\ref{Sec3}.

\begin{figure*}[htp]
\center
\includegraphics[width=7in]{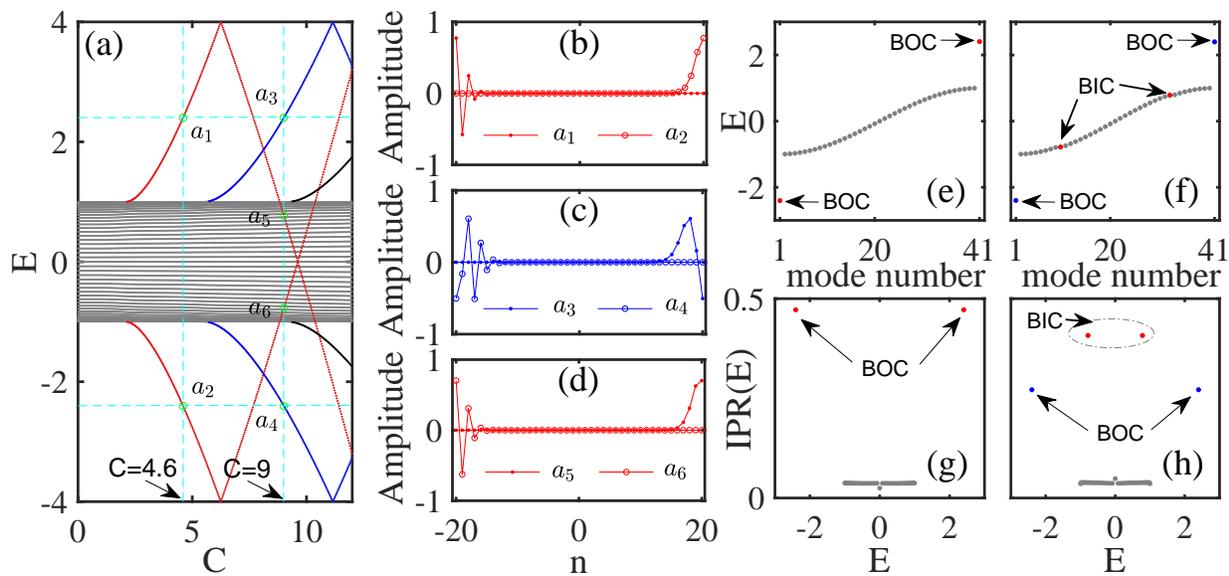}
\caption{(Color online) (a) Quasienergy $E$ versus the coupling strength $C$, where $a_{1,2}$
denote the Floquet-surface BOCs at $C=4.6$, and $a_{3,4}$ ($a_{5,6}$)  denote
the Floquet-surface BOCs (BICs) at $C=9$.
(b)--(d) The eigenstate profiles corresponding to $a_{1,2}$, $a_{3,4}$ and $a_{5,6}$ marked in (a).
(e)--(h): (top) the quasienergy spectrum and (bottom) inverse participation ratio with different coupling strengths,
$C=4.6$ for (e) and (g), and $C=9$ for (f) and (h).
The other parameters are chosen as $\xi=\omega=8$, $F=1$, and the total lattice number $N=41$.} \label{fig1}
\end{figure*}

\section{resonance between the periodic driving and tilt induced Floquet-surface BICs and BOCs \label{Sec2}}
\subsection{The model}
We consider the coherent hopping dynamics of a quantum particle in
a 1D periodically driven optical lattice subjected to a tilted potential,
which is described by the single-band tight-binding Hamiltonian
\begin{eqnarray} \label{equaion2}
H(t)&=&J(t)\sum\limits_{n}(|n\rangle\langle n+1|+|n+1\rangle\langle n|) \nonumber \\
&&-\xi\sum\limits_{n}n(|n\rangle\langle n|).
\end{eqnarray}
Here, $|n\rangle$ represents the Wannier state localized at the $n$th site ($n=0,\pm1,\pm2,...$),
and $J(t)=C+F\cos(\omega t)$ denotes the time-dependent hopping strength between adjacent sites~\cite{ma2011photon},
where $C$ is the constant hopping strength, $F$ and $\omega$ are the driving amplitude and frequency, respectively.
$\xi$ is the lattice tilt.
The model has been investigated in different physical contexts.
It describes, for example, coherent transport of ultracold atoms
in periodically-shaken optical lattices~\cite{tan2014,tan2016} and
light propagation in arrays of periodically-curved waveguides~\cite{Garanovich2012}.
Such a system also can be realized experimentally by applying an
amplitude-modulated laser standing wave and a linear potential produced by
a magnetic field gradient~\cite{ma2011photon, chen2011,Simon2011}. In the
case of resonance, i.e., for $\omega=|\xi|$, resonant interplay between the
periodic driving and tilt may lead to delocalization, which was previously
studied in Refs.~\cite{ma2011photon,tan2014,tan2016}. As we will show in our work, the resonant interplay
can induce localization at the surface of lattice, and
enables to observe surface
bound states with a quasienergy embedded in the spectrum of scattered states, which we call Floquet-surface BICs.

To do this, according to the time-dependent Schr\"{o}dinger equation $-id\psi(t)/dt=H(t)\psi(t)$
(by setting $\hbar=1$) with $\psi(t)=\sum_{n}a_n(t)|n\rangle$ and applying
the gauge transformation $a_n(t)=e^{-i\xi n t}\phi_n(t)$,
we can obtain the coupled-mode equations with probability amplitudes $\phi_n(t)$ satisfying
\begin{eqnarray} \label{equaion3}
-i\frac{d\phi_n(t)}{dt}=\Omega(t) \phi_{n-1}(t)+\Omega^*(t) \phi_{n+1}(t),
\end{eqnarray}
where, $\Omega(t)=J(t)e^{i\xi t}$ and $\Omega^*(t)$ is the complex conjugate,
which satisfies $\Omega(t)=\Omega(t+T)$ with  $T=2\pi/\omega=2\pi/|\xi|$. Then
the system can be described by an effective model without tilt.
According to the Floquet theorem, the evolution of a time-dependent
system obeys $\phi_n(t')=U(t',t)\phi_n(t)$, where $U$ is the time evolution operator
\begin{eqnarray} \label{equaion4}
U(t',t)=Q\{\exp[-i\int_t^{t'}H(t'')dt'']\},
\end{eqnarray}
where $Q$ is a chronological operator.
Then the quasienergy of the system $E$ can be obtain by diagonalizing
the Floquet Hamiltonian $H_f$, which satisfies $e^{-iH_fT}\equiv U(T,0)$.
As is well known, quasienergies
are defined apart from integer multiples of $\omega$, and conventionally
they are restricted to the interval ($-\omega/2\leq E\leq \omega/2$).
Once the quasienergies and the corresponding
eigenvectors are determined, the search of bound states, either
embedded or outside the spectrum of scattered
states, is done by inspection of the inverse participation ratio (IPR).
For the $i$th quasienergy eigenstate $\varphi(E_i)$, $1\leq i\leq N$,
which is spanned as $\varphi(E_i)=\sum_n \phi_n^i|n\rangle$ in the single-particle Hilbert space,
the IPR is defined as~\cite{kramer93, wang2017quantum}
\begin{eqnarray} \label{equaion7sz}
\mbox{IPR}(E_i)=\frac{\sum_n|\phi_n^i|^4}{(\sum_n |\phi_n^i|^2)^2}.
\end{eqnarray}
Obviously, the IPRs of the localized (bound) states have nonzero values,
and the IPRs of the extended (scattered) states are in practice zero for large $N$.
A typical example is displayed in Fig.~\ref{fig1} by choosing total lattice sites $N=41$,
$\xi=\omega=8$ and $F=1$. An inspection of the quasienergy diagram shows that, as the coupling strength
increases to $C>2.1$, Floquet-surface BOCs emerge
in pairs, above and below the band of scattered states, which are
clearly visible as isolated dispersion curves that detach from the
continuous band of scattered states, as shown in Fig.~\ref{fig1}(a).
The number of Floquet-surface BOCs always increases in pairs as the coupling strength further increases.
Especially, in the strong coupling region of $8.8<C<10.3$, the
dispersion curves of a pair of Floquet-surface BOCs that firstly penetrate into the band of scattered
states, and Floquet-surface BICs are clearly visible in the participation
ratio diagram [see Figs.~\ref{fig1}(f) and \ref{fig1}(h)].
An important property of the Floquet-surface BICs and BOCs
is that they are localized at the left and right edges of the lattice,
so that it corresponds to single-particle surface state of the Tamm type in the one-dimensional lattice,
as shown in Figs.~\ref{fig1}(b)--(d).
This explains the physical origin of Floquet-surface BICs and BOCs:
the resonance interplay between the periodic driving and tilt pushes the particle near the edges of
the lattice, which will be clarified in the next
section. It should be noted that, as opposed to single-particle
Floquet-bulk BICs recently predicted in Ref.~\cite{longhi2219},
Floquet-surface BICs are robust against parameter fluctuations and exist in a wide parametric region.
Note also that the
localizations of Floquet-surface BOCs in the same quasienergy
but different parameters [e.g., ($a_1$, $a_3$) and ($a_2$, $a_4$) in Fig.~\ref{fig1} (a)] are different, and so are the Floquet-surface BICs and BOCs in the same parameter but different quasienergies [e.g., $a_3$, $a_4$, $a_5$ and $a_6$ in Fig.~\ref{fig1} (a)]. This feature may provide a promising approach for detecting these Floquet-surface states, as discussed later.

\subsection{Localization property and robustness of Floquet-surface BICs and BOCs \label{B}}

\begin{figure}[htp]
\center
\includegraphics[width=3.6in]{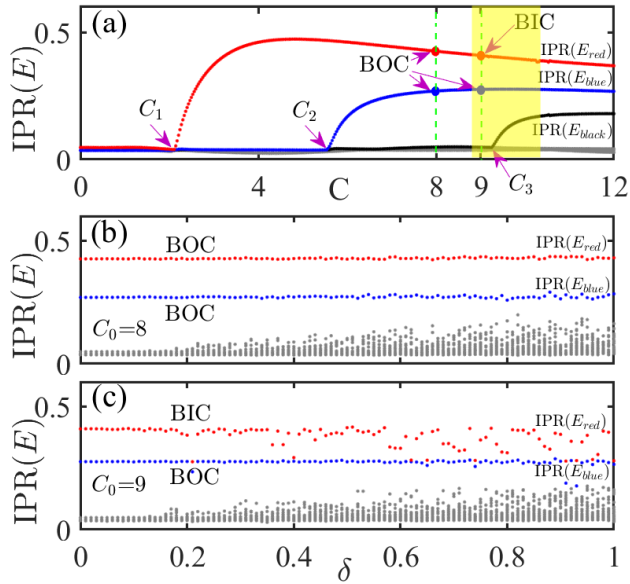}
\caption{(a) IPRs of all of Floquet-surface states as a function of the coupling strength $C$,
where the yellow area indicates the existence of Floquet-surface BICs.
The IPR values denoted by the red, blue and black lines
correspond to quasienergies shown by the red, blue and black lines in Fig.~\ref{fig1}(a), respectively.
(b) and (c) IPRs as a function of the disorder strength $\delta$
for different coupling strengths $C=8$ and $9$, respectively.
The other parameters are chosen as $\xi=\omega=8$, $F=1$ and the total lattice number $N=41$.} \label{fig2}
\end{figure}

In this subsection, we will investigate the localization property and robustness of Floquet-surface BICs and BOCs.
To study the localization property of all Floquet-surface states,
we compute the IPRs of all Floquet-surface states as a function of the coupling strength $C$,
as shown in Fig.~\ref{fig2}(a).
It clearly shows that there exist
three transition points at $C_1$, $C_2$ and $C_3$ ($C_1\simeq2.1, C_2\simeq5.6$ and $C_3\simeq9.2$),
which correspond to the appearance of new Floquet-surface states,
and the localized degrees of Floquet-surface states that primarily emerge are stronger than the later ones,
that is, $\mbox{IPR}(E_{red})>\mbox{IPR}(E_{blue})>\mbox{IPR}(E_{black})$ for a fixed coupling.
In the yellow area indicating the existence of Floquet-surface BICs,
the localized degrees of the Floquet-surface BICs are stronger than Floquet-surface BOCs.
As an example with the coupling strength $C=9$ shown in Fig.~\ref{fig1}(h),
the localized degrees of a pair of Floquet-surface BICs are obviously stronger
than that of a pair of Floquet-surface BOCs.
Because the system satisfies chiral symmetry,
the localized degrees of a symmetric pair of Floquet-surface states at the left and right edges around $0$ are same for a given coupling strength.

The Floquet-surface states emerge in an ideal finite lattice with perfectly homogeneous coupling strength.
However, the coupling strength  $C$ in reality may have non-negligible fluctuations whose effects need to be evaluated.
As shown in previous studies~\cite{molina2012,longhi2014,della2013,gallo2015,longhi2219}, lattice imperfections or disorder are expected to destroy the Floquet-surface BICs,
which decays into a resonance surface state. Generally, single-particle BICs are
fragile states,
which decay into resonance states by small perturbations.
Here, we show that the Floquet-surface BICs possess relatively stronger robustness for the parameter perturbations.
To this end, disorder is added to the constant coupling strength and yields $C=C_0+\delta \chi$,
where $C_0$ is a homogeneous coupling strength considered in the previous subsection,
$\chi$ is a random number uniformly distributed in the range $(-1,1)$,
and $\delta$ measures the strength of disorder. Figs.~\ref{fig2}(b) and \ref{fig2}(c) show typical results of IPRs versus disorder strength $\delta$ for $C_0=8$ and $C_0=9$, respectively.
The Floquet-surface BICs obviously are robust
for a disorder strength smaller than $0.2$.
Although Floquet-surface BOCs are not sensitive to the perturbation of disorder,
strong disorder can lead to the Anderson localization of bulk states (grey dots),
which may cause invisibility of the surface state for large disorder strength.
It means that Floquet-surface BICs and BOCs induced
by the resonance between the periodic driving and tilt in our system
are robust against appropriate parameter changes or fluctuations.

\subsection{Detecting the transition point of Floquet-surface states by quantum walks \label{c}}

\begin{figure}[htp]
\center
\includegraphics[width=3.6in]{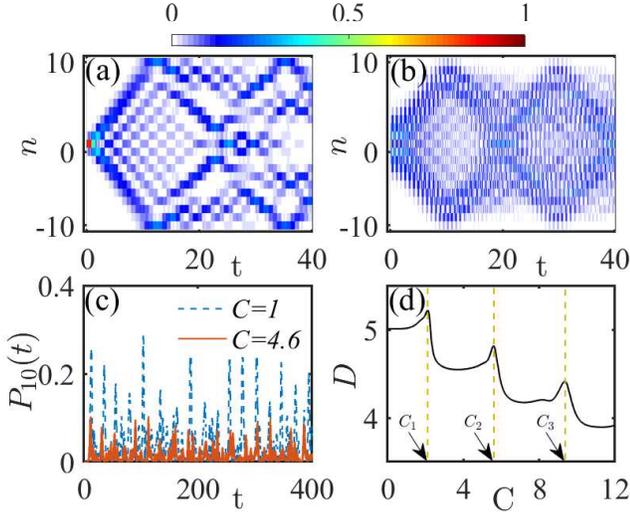} %yanhua_Tat_zz
\caption{(Color online) Repulsion effect of the Floquet-surface states.
(a) and (b) Long-time dynamical evolution for two different coupling strengths
(a) $C=1$ and (b) $C=4.6$. (c) Time-dependent probability distribution on
left edge of the lattice (n=10).
(d) Long-time average of edge degree $D$ as a function of the coupling strengths $C$. The quantum walk
is initially positioned on the center site 0.
The other parameters are chosen as $\xi=\omega=8$, $F=1$, and the total lattice number $N=21$.} \label{fig3}
\end{figure}

In this subsection, we investigate the dynamics of the quantum walks initially
located in the middle of a lattice with site number $N=21$.
As is shown in Figs.~\ref{fig3}(a) and ~\ref{fig3}(b),
the quantum walks initiated from the center
site expands ballistically and no localization phenomenon is
shown for two different hopping strengths $C=1$ (Floquet-surface state is absent) and $C=4.6$ (A pair of Floquet-surface BOCs are present).
However, close and careful
observation reveals an intriguing effect of the Floquet-surface states.
If we focus on the two boundary sites of
the lattice in Figs.~\ref{fig3}(a) and \ref{fig3}(b), the edge probability in the absence of Floquet-surface state is smaller than that in the presence of Floquet-surface state. Similar to the repulsion effect of the topologically
protected edge state shown in Ref.~\cite{wang2017quantum}, this also can be seen as a repulsion effect of the
the Floquet-surface states, and its strength is determined by the localization
properties of the Floquet-surface states. To make it clearer, we show the
time-dependent distribution on left
edge of the lattice, $P_{10}(t)$, for a long time.
It is evident that, because there is no Floquet-surface state for $C=1$,
the quantum walk can easily reach the left boundary site.
Conversely, for $C=4.6$ where Floquet-surface states exist,
the quantum walk is repelled from reaching the left boundary site
as the distribution
in the $10$th site remains a very small value all the time, see the red solid line Fig.~\ref{fig3}(c).

Interestingly, we find that the repulsion effect may provides a promising approach for experimentally
detecting the transition point of the Floquet-surface state,
and then measures the number of Floquet-surface state.
In the experiment, one can detect the long-time average of edge degree
\begin{eqnarray} \label{equaion7szz}
D=\frac{1}{T'}\sum\limits_n\int_0^{T'} |n| P_n(t)dt,
\end{eqnarray}
where $P_n(t)=|a_n(t)|^2=|\phi_n(t)|^2$, and $T'$ is the total evolution time.
Larger $D$ means more distribution in the edge sites in long-time average.
In Fig.~\ref{fig3}(d), we show the value of $D$ as a function of the coupling strength $C$ by selecting $T'=400$.
Due to the repulsion effect, the value of $D$ gradually decreases
in a stepped way with the appearance of new Floquet-surface states.
The repulsion effect can be understood from two aspects.
On one hand, the larger IPR of Floquet-surface state means more localization and hence weak repulsion.
As $C$ increases, IPRs of new Floquet-surface states become smaller and lead to stronger repulsion and smaller $D$.
On the other hands, the number of Floquet-surface states increases with $C$ in a step way. The more Floquet-surface states also lead to stronger repulsion and make the value of $D$ drop in a stepped way.
It is worth noting that there are three abnormal peaks in the variational process of $D$,
which correspond to the
transition points $C_1$, $C_2$ and $C_3$ of Floquet-surface states in Fig.~\ref{fig2}(a),
where the repulsion effect  is inversely weakened.
The main reason is that the new Floquet-surface states and the scattered states
are nearly degenerate in the vicinity of the
transition point and convert to each other. Hence, it provides a possible approach for
detecting the transition point and measuring the number of Floquet-surface states in the experiment.

\section{MULTIPLE-TIME-SCALE ASYMPTOTIC ANALYSIS \label{Sec3}}
To get deeper physical insights into the properties and
the mechanism underlying the formation of Floquet-surface states,
in this section we will develop an analytical theory for Floquet-surface states in the high-frequency
limit.
We show how the resonant interplay between periodic driving and tilt introduce
effective Tamm-type defects, and then generate Floquet-surface states in a 1D defect-free lattice.
In subsection~\ref{SubIIIA}, we develop multiple-time-scale asymptotic analysis for Floquet-surface states.
In subsection~\ref{SubIIIB}, we analytically give the asymptotic phase boundary,
which can be used to determine the generated threshold of Floquet-surface states in the high-frequency
limit.
It is worth noting that although we only analyze the high-frequency region,
our analytical results also have guiding significance
for the appearance of Floquet-surface states in the low-frequency (strong coupling) region.

%\subsection{Analytical result for the weak coupling regime}
\subsection{The resonance between periodic driving and tilt induced effective Tamm-type defects}\label{SubIIIA}
We perform a multiple-time-scale asymptotic analysis (MTSAA)
of the 1D driven and tilted finite lattice in the high-frequency
limit $C\ll\max[\omega, \sqrt{|F|\omega}]$ (see, for instance, Refs.~\cite{Garanovich2008PRL,zhu2018topological}).
To this end, we rewrite Eq.~\eqref{equaion3} as
\begin{eqnarray} \label{equation20s}
-i\frac{d\phi_n(t)}{dt}=\sum\limits_m W(t;n,m)\phi_m(t),
\end{eqnarray}
with
\begin{eqnarray} \label{equation21s}
W(t;n,m)&=&\delta_{n,m+1}\Omega(t)+\delta_{n,m-1}\Omega^*(t). \nonumber
\end{eqnarray}
Here $\delta_{n,m}$ is the Kronecker delta function. For the open boundary condition, we have $\phi_{n<-N'}\equiv 0$ and $\phi_{n>N'}\equiv 0$, in which $2N'+1=N$ is the total lattice number.
Therefore, $W(t;n,m)$ can be rewritten as
\begin{eqnarray} \label{equation22s}
W(t;n,m)&=&(1-\delta_{n,-N'})\delta_{n,m+1}\Omega(t) \nonumber \\
&&+(1-\delta_{n,N'})\delta_{n,m-1}\Omega^*(t).
\end{eqnarray}
Because the coupling $\Omega(t)$ and $\Omega^*(t)$ are periodic functions, we have $W(t;n,m)=W(t+T;n,m)$, where $T=2\pi/\omega=2\pi/\xi$.
In the high-frequency limit ($\omega \gg C$), we can introduce a small parameter $\varepsilon$, which satisfies $T=O(\varepsilon)$.
Thus, the solution of Eq.~\eqref{equation20s} can be given by the series expansion
\begin{eqnarray} \label{equation23s}
\phi_n(t)&=&U_n(t_0,t_1,t_2,...)+\varepsilon v_n(t_{-1},t_0,t_1,t_2,...) \nonumber \\
&&+\varepsilon^2 w_n(t_{-1},t_0,t_1,t_2,...) \nonumber \\
&&+\varepsilon^3 \zeta_n(t_{-1},t_0,t_1,t_2,...)+O(\varepsilon^4),
\end{eqnarray}
where $t_{l'}=\varepsilon^{l'} t$.
Then the differential is performed according to the convention:
\begin{eqnarray} \label{equation24s}
\frac{d}{dt}=\varepsilon^{-1}\frac{\partial}{\partial t_{-1}}+\frac{\partial}{\partial t_0}+
\varepsilon\frac{\partial}{\partial t_1}+\varepsilon^2\frac{\partial}{\partial t_2}+\cdots.
\end{eqnarray}
In the series solution, the function $U_n$ describes the averaged behavior
\begin{eqnarray} \label{equation25s}
\langle \phi_n \rangle=U_n; \
\langle \frac{d\phi_n}{dt} \rangle=\frac{dU_n}{dt},
\end{eqnarray}
in which the average notation is given by
\begin{eqnarray} \label{equation26s}
\langle \bullet \rangle =\varepsilon T^{-1}\int_{\varepsilon^{-1} t}^{\varepsilon^{-1} (t+T)}(\bullet)(t_{-1})dt_{-1}. \nonumber
\end{eqnarray}
It is worth to note that $U_n$ does not depend on the `fast' variable $t_{-1}$, which means that
\begin{eqnarray} \label{equation27s}
\langle U_n \rangle=U_n ;
\langle \frac{dU_n}{dt} \rangle=\frac{dU_n}{dt}.
\end{eqnarray}
From Eqs.~\eqref{equation25s} and \eqref{equation27s}, we have
\begin{eqnarray} \label{equation28s}
&&\langle v_n \rangle=\langle w_n \rangle=\langle \zeta_n \rangle \equiv0 ; \nonumber \\
&&\langle \frac{\partial v_n}{\partial t_{l'}} \rangle=\langle \frac{\partial w_n}{\partial t_{l'}} \rangle=\langle \frac{\partial \zeta_n}{\partial t_{l'}} \rangle \equiv0,
\end{eqnarray}
for $l'=-1,0,1,2,\cdots$.

Substituting Eq.~\eqref{equation23s} into Eq.~\eqref{equation20s} and collecting terms with different orders of $\varepsilon$,
we can obtain a closed-form equation for $U_n$
\begin{eqnarray} \label{equation44s}
-i\frac{d U_n}{dt}=\sum\limits_{m} W_s(n,m) U_m .
\end{eqnarray}
Here the effective coupling coefficients are given by
\begin{eqnarray} \label{equation45s}
W_s(n,m)&=&W_0(n,m)+\sum\limits_j W_1(n,j,m) \nonumber \\
&&+\sum\limits_{q,j}W_2(n,q,j,m),
\end{eqnarray}
with
\begin{eqnarray} \label{}
W_0(n,m)&=&\langle W(t;n,m) \rangle \nonumber \\
&=&(1-\delta_{n,-N'})\delta_{n,m+1}\frac{F}{2} \nonumber \\
&&+(1-\delta_{n,N'})\delta_{n,m-1}\frac{F}{2}, \nonumber
\end{eqnarray}
\begin{eqnarray} \label{}
\sum\limits_j W_1(n,j,m)&=&i\sum\limits_j \langle W(t;n,j)M(t;j,m) \rangle \nonumber \\
&=&-\delta_{n,-N'}\Delta+\delta_{n,N'}\Delta, \nonumber
\end{eqnarray}
\begin{eqnarray} \label{}
\sum\limits_{q,j} W_2(n,q,j,m)&=&\sum\limits_{q,j} \langle M(t;n,q)[W(t;q,j) \nonumber \\
&&-W_0(q,j)]M(t;j,m) \rangle \nonumber \\
&&+\sum\limits_{q,j} \langle M(t;n,q)[W_0(q,j)M(t;j,m) \nonumber \\
&&-M(t;q,j)W_0(j,m)] \rangle \approx 0, \nonumber
\end{eqnarray}
where $M(t;n,m)=\int_0^{t}[W(t';n,m)-W_0(n,m)]dt'$.
Finally the effective equations for the slowly varying functions $U_n$ read as
\begin{eqnarray} \label{equaion5sa}
-i\frac{dU_n}{dt}&=&\frac{F}{2} U_{n+1}+\frac{F}{2} U_{n-1}
-\delta_{n,-N'}\Delta U_{-N'} \nonumber \\
&&+\delta_{n,N'}\Delta U_{N'}.
\end{eqnarray}
Here the effective energy bias $\Delta=\omega^{-1}(C^2+F^2/8)$,
which describes the virtual
defects at boundaries, as shown in the schematic diagram in Fig.~\ref{fig4}(a).

Based on the above discussions, the periodically driven and tilted
system can be described by effective static coupled mode
equations~\eqref{equaion5sa} without tilt. The major difference is the existence of virtual Tamm-type
defects at boundaries in the effective model. Similar to a surface
perturbation, the virtual defects can form defect-free surface
states~\cite{Garanovich2008PRL,zhu2018topological}.
Therefore, in our system, without any embedded or
nonlinearity-induced defects, the surface perturbation (virtual
defect) is induced by the resonant interplay between periodic driving and tilt,
which is the primary reason of appearing Floquet-surface states.
In the next subsection, we will give the parameter regions of Floquet-surface states.

\subsection{Asymptotic phase boundary and phase diagram} \label{SubIIIB}
To estimate the cutoff values (phase boundaries) for the
regions of Floquet-surface states caused by virtual defects,
We will study the phase diagram of Floquet-surface states about coupling strength $C$ and driving amplitude $F$.
We consider stationary solutions in the form of $U_n(t)=U_n(0)e^{iEt}$ with $E$ being quasienergy.
Substituting it into Eq.~\eqref{equaion5sa}, we obtain
\begin{eqnarray} \label{equation49s}
E U_{n}&=&\frac{F}{2} U_{n+1}+\frac{F}{2} U_{n-1}-\delta_{n,-N'}\Delta U_{-N'} \nonumber \\
&&+\delta_{n,N'}\Delta U_{N'}.
\end{eqnarray}
For an infinite lattice, we have
\begin{eqnarray} \label{equation50s}
E_b U_{n}=\frac{F}{2} U_{n+1}+\frac{F}{2} U_{n-1}.
\end{eqnarray}
The solution of Eq.~\eqref{equation50s} can be given by the ansatz
\begin{eqnarray} \label{equation51s}
U_{n}=Ye^{ikn}+Ze^{-ikn},
\end{eqnarray}
where $Y$ and $Z$ are undetermined coefficients.
Substituting Eq.~\eqref{equation51s} into Eq.~\eqref{equation50s},
we can obtain the band of scattered states $E_b=F\cos(k)$ with $k\in[-\pi,\pi]$.

\begin{figure}[!htp]
	\center
	\includegraphics[width=3.6in]{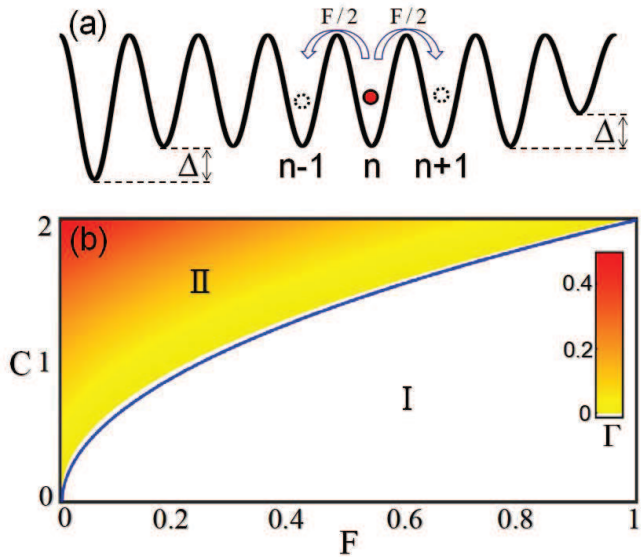}%{Fig3_zsszz.eps}
	\caption{(Color online) (a) Schematic diagram for the effective model Eq.~\eqref{equaion5sa}.
		(b) Phase diagram of Floquet-surface states, where the colors denote the value of gap parameter $\Gamma$.
		The white region (I) does not support Floquet-surface states,
		and the colorized region (II) supports Floquet-surface states.
		The blue curve corresponds to the phase boundary, which satisfies Eq.(\ref{equaion7}).
		The other parameters are chosen as $\xi=\omega=8$, and the total lattice number $N=41$.} \label{fig4}
\end{figure}

For a finite lattice with sufficiently large number of sites, considering the two edges, we have
\begin{eqnarray} \label{equation54s}
EU_{-N'+1}&=&\frac{F}{2}U_{-N'}+\frac{F}{2}U_{-N'+2}, \nonumber \\
EU_{-N'}&=&\frac{F}{2}U_{-N'+1}-\Delta U_{-N'}, \nonumber \\
EU_{N'}&=&\frac{F}{2}U_{N'-1}+\Delta U_{N'}, \nonumber \\
EU_{N'-1}&=&\frac{F}{2}U_{N'}+\frac{F}{2}U_{N'-2}.
\end{eqnarray}
Besides $U_{-N'}$ and $U_{N'}$, the coupling equations are consistent with Eq.~\eqref{equation50s}, so that we should rewrite the ansatz similar to Eq.~\eqref{equation51s}, i.e.,
\begin{eqnarray} \label{equation51a}
U_{n}&=&U_n \qquad (n=-N', N'), \\
U_{n}&=&Ye^{ik(n+N')}+Ze^{-ik(n+N')} \quad (-N'<n< N'). \nonumber
\end{eqnarray}
First, we consider left boundary of the lattice and we can give a set of equations
\begin{eqnarray} \label{equation51b}
%EU_{-N'+1}&=&\frac{F}{2}U_{-N'}+\frac{F}{2}U_{-N'+2}, \nonumber \\
EU_{-N'}&=&\frac{F}{2}U_{-N'+1}-\Delta U_{-N'}, \nonumber \\
EU_{-N'+n'}&=&\frac{F}{2}U_{-N'+n'-1}+\frac{F}{2}U_{-N'+n'+1},
\end{eqnarray}
where $n'>0$, combining Eq.~\eqref{equation51a} and Eq.~\eqref{equation51b}, we have
\begin{eqnarray} \label{equation51c}
e^{-i2kn'}=\frac{-Fe^{ik}+2\Delta}{-Fe^{-ik}+2\Delta}.
\end{eqnarray}
We set $k=-i\varrho$ and have $e^{-i2kn'}=e^{-2 \varrho n'}$, where $\varrho$ is real number. If $\varrho>0$, when $n'\rightarrow \infty$, we have $e^{-2 \varrho n'}\simeq 0$ and $-Fe^{\varrho}+2\Delta=0$.
If $\varrho<0$, when $n'\rightarrow \infty$, we have $e^{-2 \varrho n'}\simeq \infty$ and $-Fe^{-\varrho}+2\Delta=0$.
Consequently, $e^{\pm\varrho}$ are given by
\begin{eqnarray} \label{equation55s}
&&e^{\varrho}=\frac{2\Delta}{F}=d, \nonumber \\
&&e^{-\varrho}=\frac{F}{2\Delta}=d^{-1}.
\end{eqnarray}
Thus the left Floquet-surface state induced by the effective
Tamm-type defect with the quasienergy $E_{s1}$ is given by
\begin{eqnarray} \label{equation56}
E_{s1}=\frac{F}{2}(e^{ik}+e^{-ik})=\frac{F}{2}(d+d^{-1})=\Delta+\frac{F^2}{4\Delta}.
\end{eqnarray}
When we consider the right boundary of the lattice, we can also obtain the quasienergy
of the right Floquet-surface $E_{s2}=-E_{s1}$.
Obviously, when $|E_{s1}|>\max(E_b)$, there exist Floquet-surface states.
$\max(E_b)$ is given by $\cos(k)=1$.
Then one can obtain the cutoff value
\begin{eqnarray} \label{equaion7}
C=\sqrt{\frac{ F}{2}(\omega-\frac{F}{4})}.
\end{eqnarray}
The cutoff value defines the boundary between the regions with and without Floquet surface
states, see the blue curve in Fig. \ref{fig4}(b).
Especially, for a fixed $C$, the region of existing Floquet-surface
states can be tuned by the driving amplitude.

To verify the above analytical results, we numerically calculate
the quasienergy spectra under open boundary condition.
Combining the band of scattered states $E_b$ and quasienergy $E$,
we define a parameter,
\begin{equation}
\Gamma=\max(E)-\max(E_b),
\end{equation}
which represents the energy gap between Floquet-surface states and the band of scattered states. $\Gamma=0$ indicates the absence of Floquet-surface states. Otherwise, $\Gamma>0$ indicates the appearance of Floquet-surface states.
In Fig.~\ref{fig4}(b), we numerically show the phase
diagram of Floquet-surface states in the parameter plane $(C,F)$, where the colors denote the gap parameter $\Gamma$.
The white region does not support Floquet-surface states,
and the colorized region supports Floquet-surface states.
Our numerical results clearly show that the phase boundary well agrees with our analytical result; see the blue solid line in Fig.~\ref{fig4}(b).

\section{Conclusion \label{Sec5}}
In summary, We have studied the Floquet-surface states
in a resonantly driven 1D tilted defect-free lattice.
It is found that the Floquet-surface BICs and BOCs can be induced in such a system by using the
resonant interplay between the periodic driving and tilt.
Compared with single-particle Floquet-bulk BICs~\cite{longhi2219},
which are fragile states
and whose existence requires fulfillment of certain
condition, Floquet-surface BICs can exist in a wide range of parameter space
and are structurally stable against perturbations of system parameter.
Analytical results are derived in the high-frequency limit by a multiple-time-scale asymptotic analysis.
It is found that the resonance between the periodic driving and tilt can induce effective Tamm-type defects
at boundaries of the lattice,
and thus results in the appearance of Floquet-surface states.
According to the asymptotic analysis,
the phase boundary of existing Floquet-surface states is analytically given,
$C=\sqrt{{ F}/{2}(\omega-{F}/{4})}$.
The region of existing Floquet-surface states
can be adjusted by tuning the coupling strength or the driving amplitude.

With currently available techniques, it is possible
to realize our model and observe our theoretical predictions
with experiments. Our proposed titled lattices with resonantly driven can be demonstrated experimentally
in numerous cold-atom setups~\cite{aide2011,aide2013,aides2013,miya2013,Kennedy2013,aide2015}.
For instance, one can use Wannier-Stark ladder with large static energy offset $\xi$
\cite{aide2013,miya2013,Dimitrova2020} to realize a 1D tilted lattice.
Periodic driving can be introduced by harmonically modulating the tunneling rate at the tilted frequency
$\xi$~\cite{ma2011photon}.
For such
a resonantly driven 1D tilted lattice,
Floquet-surface BICs and BOCs may be observed via Quantum walks.
Our work paves a way to the experimental realization of BICs in a single-particle quantum system.

\begin{acknowledgments}
The authors are thankful to Chaohong Lee for enlightening suggestions and helpful discussions.
This work is supported by the National Natural Science Foundation of China under
Grant No. 11805283 and No. 11874434, the Hunan Provincial Natural
Science Foundation under Grants No. 2019JJ30044,
the Scientific Research Fund of Hunan Provincial Education Department
under Grant No. 19A510, and the Talent project of Central South University
of Forestry and Technology under Grant No. 2017YJ035.
Y.K. is partially supported by the Office of China Postdoctoral Council (Grant No. 20180052), the National Natural Science Foundation of China (Grant No. 11904419), and the Australian Research Council (DP200101168).

\end{acknowledgments}

\bibliography{tilted_lattice}

\end{document}